\documentclass[pdflatex,sn-mathphys-num]{sn-jnl}
\usepackage{graphicx}%
\usepackage{multirow}%
\usepackage{amsmath,amssymb,amsfonts}%
\usepackage{amsthm}%
\usepackage{mathrsfs}%
\usepackage[title]{appendix}%
\usepackage{xcolor}%
\usepackage{textcomp}%
\usepackage{manyfoot}%
\usepackage{booktabs}%
\usepackage{algorithm}%
\usepackage{algorithmicx}%
\usepackage{algpseudocode}%
\usepackage{listings}%

\theoremstyle{thmstyleone}%

\theoremstyle{thmstyletwo}%

\theoremstyle{thmstylethree}%
\newcommand\gbf{\mathbf{g}}

\newcommand\ubf{\mathbf{u}}

\newcommand\xbf{\mathbf{x}}

\newcommand\Gbf{\mathbf{G}}

\newcommand\Pbf{\mathbf{P}}
\newcommand\Qbf{\mathbf{Q}}

\newcommand\Ubf{\mathbf{U}}

\newcommand\Wbf{\mathbf{W}}
\newcommand\Xbf{\mathbf{X}}

\newcommand{\rhob}{\boldsymbol{\rho}}

\allowdisplaybreaks

\begin{document}

\title[]{Neural Constraints on Cognitive Experience and Mental Health}


\author[1]{\fnm{Bita} \sur{Shariatpanahi}}
\equalcont{These authors contributed equally to this work.}
\author[2,3,4,5]{\fnm{Erfan} \sur{Nozari}}
\equalcont{These authors contributed equally to this work.}
\author[1]{\fnm{Soroush} \sur{Daftarian}}
\author[3]{\fnm{Fahimeh} \sur{Arab}}
\author[6]{\fnm{Mina} \sur{Kheirkhah}}
\author[1]{\fnm{Felix P.} \sur{Bernhard}}
\author[1]{\fnm{Shiva} \sur{Khodadadi}}
\author[7,8]{\fnm{Erik J.} \sur{Giltay}}
\author[7]{\fnm{Kaat} \sur{Hebbrecht }}
\author[9]{\fnm{Stefan} \sur{G. Hofmann}}
\author[10]{\fnm{Tim} \sur{Hahn}}
\author*[1,11]{\fnm{Hamidreza} \sur{Jamalabadi}}\email{hamidreza.jamalabadi@uni-marburg.de}

\affil[1]{\orgdiv{Department of Psychiatry and Psychotherapy}, \orgname{University of Marburg}, \country{Germany}}
\affil[2]{\orgdiv{Department of Mechanical Engineering}, \orgname{University of California}, \country{Riverside, CA, United States}}
\affil[3]{\orgdiv{Department of Electrical and Computer Engineering}, \orgname{University of California, Riverside}, \country{Riverside, CA, United States}}
\affil[4]{\orgdiv{Department of Bioengineering}, \orgname{University of California, Riverside}, \country{Riverside, CA, United States}}
\affil[5]{\orgdiv{Neuroscience Graduate Program}, \orgname{University of California, Riverside}, \country{Riverside, CA, United States}}
\affil[6]{\orgdiv{Experimental Therapeutics and Pathophysiology Branc}, \orgname{National Institute of Mental Health}, \country{Bethesda, MD, USA}}
\affil[7]{\orgdiv{Collaborative Antwerp Psychiatric Research Institute (CAPRI)}, \orgname{University of Antwerp}, \country{Belgium}}
\affil[8]{\orgdiv{Department of Psychiatry}, \orgname{Leiden University Medical Center}, \country{The Netherlands}}
\affil[9]{\orgdiv{Department of Psychology}, \orgname{University of Marburg}, \country{Germany}}
\affil[10]{\orgdiv{Institute for Translational Psychiatry}, \orgname{University of Münster}, \country{Germany}}
\affil[11]{\orgdiv{Center for Mind, Brain, and Behavior (CMBB)}, \orgname{University of Marburg}, \country{Germany}}

\newcommand\tbd[1]{{\color{red}#1}}


\abstract{Understanding how neural dynamics shape cognitive experiences remains a central challenge in neuroscience and psychiatry. Here, we present a novel framework leveraging state-to-output controllability from dynamical systems theory to model the interplay between cognitive perturbations, neural activity, and subjective experience. We demonstrate that large-scale fMRI signals are constrained to low-dimensional manifolds, where affective and cognitive states are naturally organized. Furthermore, we provide a theoretically robust method to estimate the controllability Gramian from steady-state neural responses, offering a direct measure of the energy required to steer cognitive outcomes. In five healthy participants viewing 2,185 emotionally evocative short videos, our analyses reveal a strong alignment between neural activations and affective ratings, with an average correlation of \( r \approx 0.7 \). In a clinical cohort of 255 patients with major depressive disorder, biweekly Hamilton Rating Scale trajectories over 11 weeks significantly mapped onto these manifolds, explaining approximately 20\% more variance than chance (\( p < 10^{-10} \), numerically better than chance in 93\% reaching statistical significance in one-third of subjects). Our work bridges dynamical systems theory and clinical neuroscience, providing a principled approach to optimize mental health treatments by targeting the most efficient neural pathways for cognitive change.

}


\maketitle

\section{Introduction}
\label{sec:intro}

Why do we feel the way we do, and why is it sometimes easy—but other times difficult, if not impossible—to change our “cognitive experience”? In the clinical domain, a parallel question emerges: Why do some patients with affective disorders, such as major depressive disorder (MDD), recover quickly, while at least 50\% either fail to achieve remission or relapse rapidly despite treatment~\cite{cuijpers2024outcomes}? These questions, which probe the connection between neural dynamics and subjective experience, drive the core inquiry of this paper. 

We propose that the answers lie in the properties of the alignment between cognitive and neural representations—a phenomenon that shapes cognitive experiences into structured manifolds within the brain’s high-dimensional neural spaces~\cite{conwell2025perceptual,mcintosh2019hidden,busch2023multi}. These properties—might—reveal why cognitive states differ in their adaptability and why clinical outcomes vary so widely, offering a mathematically reachable framework to understanding them. Our argument rests on three interconnected insights (see Figure~\ref{fig:Fig1}), each showing how these properties, rooted in neural dynamics, can be precisely quantified to explain and predict cognitive outcomes:

\begin{figure}[h]
    \centering
    \includegraphics[width=\textwidth, keepaspectratio]{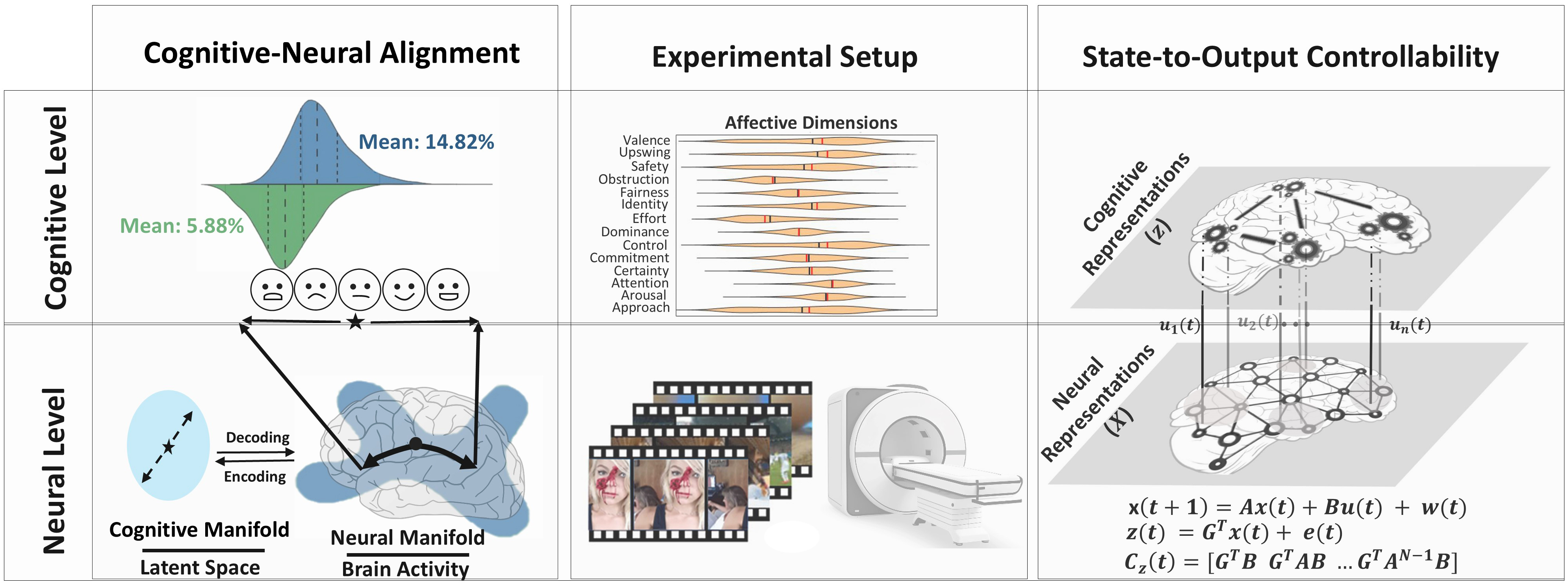}
    \caption{\textbf{Neural--Cognitive Alignment and State-to-Output Controllability Framework.} 
    \textbf{Left (Cognitive--Neural Alignment):} Cognitive experiences (e.g., emotional dimensions) map onto latent neural manifolds, capturing how large-scale brain activity encodes cognitive processes. Moving across this manifold directly translates to a new cognitive experience or, in the case of machines, newly generated stimuli that are very close to the original but have different cognitive scores (see e.g., \cite{goetschalckx2019ganalyze, rothermelcognitive}).  
    \textbf{Center (Experimental Setup):} Participants underwent fMRI while viewing a wide range of emotionally evocative video clips ~\cite{horikawa2020neural}, rated along multiple affective and cognitive dimensions. These stimuli serve as “perturbations,” allowing estimation of brain responses across diverse cognitive states.
    \textbf{Right (State-to-Output Controllability):} By modeling neural states ($x$) and cognitive outputs ($z$) in a linear system, we assess if and how cognitive perturbations can shift brain dynamics and thus alter cognitive outcomes. The matrix equations ($A, B, G$) represent the system’s dynamics and output mapping. See Section~\ref{sec:theory} for details.}
    \label{fig:Fig1}
\end{figure}

First, cognitive experiences are shaped—and constrained—by the geometry of neural manifolds. Although neural activity spans a vast and complex landscape, it is governed by structural and functional constraints that determine which states can be reached and which transitions are possible~\cite{medaglia2017brain,gu2015controllability,lynn2019physics}. Picture a cognitive experience, like the emotional tone of a memory, as a point on a curved surface: moving to a nearby point (e.g., from neutral to slightly positive) might be easy, while climbing to a distant or opposing state (e.g., reversing a deep sadness) could be steep or obstructed. Network control theory has explored these limits in neural terms~\cite{lydon2021modeling,jamalabadi2021missing,hahn2023towards}, but we argue that the manifold itself serves as a geometric scaffold for cognition, dictating how experiences evolve and why some changes are inherently harder.

Second, these neural dynamics can be captured effectively with linear models at the macroscale. Despite the brain’s intricate complexity, our recent analysis of EEG and fMRI data~\cite{nozari2024macroscopic} shows that linear dynamical models consistently outperform nonlinear alternatives, excelling in predictive accuracy and computational simplicity while leaving little unexplained. This finding echoes decades of research—from Dynamic Causal Modeling (DCM)\cite{friston2003dynamic} to system-identification techniques\cite{sani2021modeling}—where linear formulations reliably describe robust neural patterns. For cognitive experiences, this linearity means that shifts along the manifold, such as a change in emotional arousal or valence, can be approximated - at least locally -  as straight-line paths, making them easier to predict and potentially manipulate.

Third, cognitive and neural representations in large encoder-decoder models, such as VAEs, GANs, and diffusion models, align remarkably well, with properties like subjective aesthetics, emotional arousal, and valence emerging as linear gradients in their latent spaces~\cite{goetschalckx2019ganalyze,goetschalckx2021generative} (see Figure \ref{fig:Fig1}). Accumulating evidence suggests this alignment mirrors how the human brain operates, indicating a deep parallel between artificial and biological systems~\cite{goldstein2024alignment,kanwisher2023using}. Our recent work~\cite{rothermelcognitive} exemplifies this, showing that the latent space of a Very Deep Variational Autoencoder (VDVAE) maps tightly to fMRI brain responses from subjects viewing thousands of images. When the VDVAE’s latent space is perturbed—say, shifting an image’s valence from neutral to positive—the fMRI-reconstructed images reflect the same change; similarly, altering the brain data produces matching shifts in the machine’s output, and these effects are robust across subjects. Thus, cognitive and neural representations align so closely that cognitive properties consistently appear as linear gradients, both in artificial latent spaces and the brain’s neural responses. This alignment provides a common mathematical framework for modeling and controlling cognitive experiences.

Based on these insights, we formalize our central questions mathematically—i.e., which experiences are possible and how they unfold over time—as a state-to-output controllability problem~\cite{danhane2023characterizations}. Specifically: can a new experience (hereafter referred to as a “cognitive perturbation”) effectively alter neural states to achieve desired cognitive outcomes?
In our formulation, an external cognitive stimulus acts simultaneously as an input ("cognitive perturbation") and generates an output ("cognitive experience"), "filtered" by neural dynamics (see Figure~\ref{fig:Fig1}).

In what follows, we provide a formal description of this framework and formulate two testable hypotheses. We then test these hypotheses on (i) a large-scale neuroimaging dataset (Section~\ref{sec:fMRIAC}), and (ii) clinical data from a large cohort of patients with depression who received inpatient treatment (Section~\ref{sec:validation}).

\section{Theoretical Framework and Results}
\label{sec:theory}

To understand how neural dynamics shape cognitive experiences and why some transitions are easier than others, we begin with a linear formulation that integrates all key components: neural states, cognitive perturbations, and their projections onto cognitively relevant gradients. Following our recent findings~\cite{nozari2024macroscopic,rothermelcognitive}, we model these neural-cognitive dynamics as:
\begin{subequations}\label{eq:xz}
\begin{align}
    \mathbf{x}(t+1) &= \mathbf{A}\,\mathbf{x}(t) + \mathbf{B}\,\mathbf{u}(t), \\
    \mathbf{z}(t) &= \mathbf{G}^\top\,\mathbf{x}(t).
\end{align}
\end{subequations}
Here, \(\mathbf{x}(t) \in \mathbb{R}^N\) is the high-dimensional vector of neural states (e.g. whole-brain fMRI voxel activations), capturing the brain’s full activity at time \(t\). The vector \(\mathbf{u}(t) \in \mathbb{R}^n\), where \(n \ll N\), represents low-dimensional cognitive perturbations (e.g., emotional ratings of stimuli like videos), acting as inputs that nudge the neural system. The output \(\mathbf{z}(t) \in \mathbb{R}^n\) is the projection of \(\mathbf{x}(t)\) onto a neural manifold tied to cognitive experience—still neural in nature, but its reduced size reflects movement along gradients (e.g., valence or arousal) critical for generating subjective states. The matrices \(\mathbf{A} \in \mathbb{R}^{N \times N}\), \(\mathbf{B} \in \mathbb{R}^{N \times n}\), and \(\mathbf{G} \in \mathbb{R}^{N \times n}\) define the state transitions, perturbation mapping, and projection into this cognitive subspace, respectively.

This formulation allows us to address the paper’s central question—why are some cognitive experiences easy to achieve while others resist change? —within the framework of state-to-output controllability~\cite{danhane2023characterizations}. Specifically, we ask: can a cognitive perturbation \(\mathbf{u}(t)\) (e.g. watching a video or undergoing psychotherapy) steer the neural state \(\mathbf{x}(t)\) to produce any desired cognitive outcome \(\mathbf{z}(T)\) over time \(T\)? In control theory, this is answered by the state-to-output controllability Gramian:
\begin{align}\label{eq:Wz}
    \mathbf{W}_z = \sum_{\ell = 0}^{\infty} \mathbf{G}^\top \mathbf{A}^\ell \mathbf{B}\,\mathbf{B}^\top (\mathbf{A}^\top)^\ell \mathbf{G},
\end{align}
which is positive-definite if the system is controllable~\cite{bertram1960optimal}. The energy required to shift \(\mathbf{z}(t)\) by \(\Delta \mathbf{z}(T)\) is then \(\Delta \mathbf{z}(T)^\top \mathbf{W}_z^{-1} \Delta \mathbf{z}(T)\), with \(\mathbf{W}_z\)’s eigenvalues indicating the effort needed for different directions of change.

A major challenge, however, is that computing \(\mathbf{W}_z\) directly requires full knowledge of \(\mathbf{A}\), \(\mathbf{B}\), and \(\mathbf{G}\)—matrices that are impractical to estimate from empirical brain data due to their high dimensionality and complexity. To overcome this, we propose a novel theoretical approach: estimating \(\mathbf{W}_z\) using only steady-state neural responses, bypassing the need to model \(\mathbf{A}\) and \(\mathbf{B}\) explicitly. Assume we have steady-state fMRI responses \(\mathbf{x}_k\) for \(K\) distinct cognitive stimuli \(\mathbf{u}_k\) (where \(K \gg n\)), collected as:
\[
\mathbf{X} = \begin{bmatrix} \mathbf{x}_1 & \mathbf{x}_2 & \cdots & \mathbf{x}_K \end{bmatrix}, \quad \mathbf{U} = \begin{bmatrix} \mathbf{u}_1 & \mathbf{u}_2 & \cdots & \mathbf{u}_K \end{bmatrix}.
\]
From steady-state dynamics (\(\mathbf{x}_k = (\mathbf{I} - \mathbf{A})^{-1} \mathbf{B} \mathbf{u}_k\)), we define \(\mathbf{K} = \mathbf{X} \mathbf{U}^\dagger\) (where \(\dagger\) is the Moore-Penrose pseudoinverse) and estimate \(\mathbf{G}\)’s columns as fMRI gradients:
\[
\mathbf{g}_i = \mathbb{E}[\mathbf{x} \mid u_i = 1] - \mathbb{E}[\mathbf{x} \mid u_i = 0],
\]
reflecting the neural shift for each cognitive dimension \(u_i\) (normalized to [0, 1])~\cite{rothermelcognitive}. The empirical state-to-output controllability Gramian is then:
\begin{subequations}\label{eq:empirical_gramian}
\begin{align}
    \hat{\mathbf{W}}_z &= \mathbf{G}^\top \mathbf{K} \mathbf{K}^\top \mathbf{G}.
\end{align}
\end{subequations}

This approximation, detailed in Appendix~\ref{Proof_theory1}, relies solely on \(\mathbf{U}\) (stimulus ratings) and \(\mathbf{X}\) (fMRI responses), which our data—thousands of stimulus-response pairs from emotionally evocative videos (Appendix~\ref{fMRI data})—readily provides. For instance, our fMRI dataset includes 2,185 videos rated across 14 affective dimensions, paired with whole-brain voxel responses from five healthy subjects~\cite{cowen2017self}, offering ample steady-state samples to compute \(\hat{\mathbf{W}}_z\).

\begin{figure}[t]
    \centering
    \includegraphics[width=0.5\textwidth]{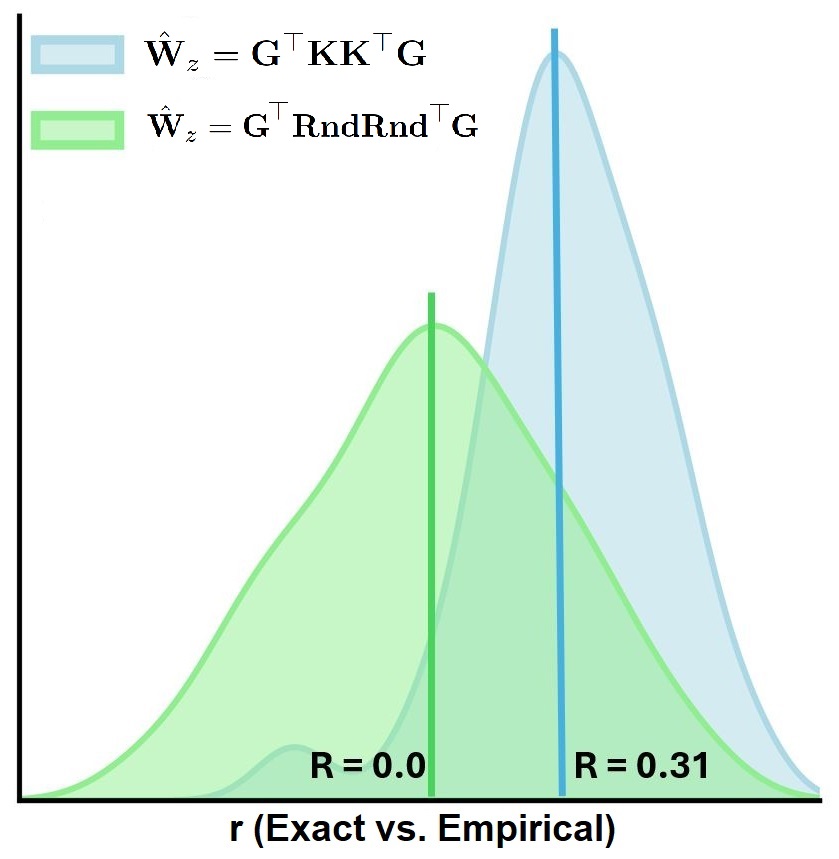}
    \caption{\textbf{Numerical Evaluation of Spectral Similarity Between Exact and Empirical State-to-Output Controllability Gramians.} We simulated 100 linear time-invariant (LTI) systems (Eq.~\eqref{eq:xz}) with \(N = 1000\) and \(n = 14\), using known \(\mathbf{A}\) and \(\mathbf{B}\) matrices to generate neural dynamics. Each system underwent 200 trials with constant cognitive perturbations \(\mathbf{u}_k(t)\) (\(k = 1, \dots, 200\)), from which \(\mathbf{X}\) and \(\mathbf{U}\) were derived, and \(\mathbf{G}\)’s gradients were estimated. The exact \(\mathbf{W}_z\) was computed using Eq.~\eqref{eq:Wz} with ground-truth \(\mathbf{A}\) and \(\mathbf{B}\), while the empirical \(\hat{\mathbf{W}}_z\) used Eq.~\eqref{eq:empirical_gramian}. The Hungarian algorithm ensured eigenvalue ordering~\cite{kuhn1955hungarian}. The figure displays the distribution of mean correlation coefficients \(r\) (exact vs. empirical), with the empirical approach (blue, mean \(r = 0.31\)) showing significantly higher similarity than chance (green, centered at \(r = 0.0\)), supporting the validity of our steady-state estimation method.}
    \label{fig:Fig2}
\end{figure}

To validate this theoretical approach, we conducted simulations to compare the exact and empirical Gramians. We generated data from 100 linear time-invariant (LTI) systems (Eq.~\eqref{eq:xz}) with \(N = 1000\) and \(n = 14\), using known \(\mathbf{A}\) and \(\mathbf{B}\) matrices to simulate neural dynamics. Each system underwent 200 trials with constant cognitive perturbations \(\mathbf{u}_k(t)\) (\(k = 1, \dots, 200\)), mimicking diverse stimuli. From these, we derived \(\mathbf{X}\) and \(\mathbf{U}\) matrices and estimated \(\mathbf{G}\)’s gradients, then computed the exact \(\mathbf{W}_z\) using Eq.~\eqref{eq:Wz} with ground-truth \(\mathbf{A}\) and \(\mathbf{B}\), and the empirical \(\hat{\mathbf{W}}_z\) using Eq.~\eqref{eq:empirical_gramian}. The Hungarian algorithm ensured proper eigenvalue ordering across runs~\cite{kuhn1955hungarian}. Figure~\ref{fig:Fig2} shows the distribution of mean correlation coefficients \(r\) (exact vs. empirical), demonstrating that the empirical method (blue) achieves a mean \(r = 0.31\), significantly higher than chance (green, centered at \(r = 0.0\)), validating our steady-state estimation approach.

This framework yields two testable hypotheses:
\begin{enumerate}
    \item \textbf{Structural Alignment}: The linear model predicts that the correlation structure of cognitive inputs (\(\mathbf{U}\)) mirrors that of neural projections (\(\mathbf{G}\)), testable by comparing eigenvalues and eigenvectors of their partial correlation matrices (Appendix~\ref{Proof_theory2}).
    \item \textbf{Trajectory Predictability}: If macroscale dynamics are approximately linear, \(\hat{\mathbf{W}}_z\)’s eigenvalues and eigenvectors should align with observed neurocognitive trajectories, where large eigenvalues indicate easily navigable directions and small ones signal high-effort transitions.
\end{enumerate}

These hypotheses, tested in Sections~\ref{sec:fMRIAC} and~\ref{sec:validation}, leverage our steady-state approach to reveal how neural constraints shape cognitive and clinical outcomes.

\section{The Parity of Cognitive and Neural Manifolds}
\label{sec:fMRIAC}

We first investigate whether the system described by Eq.~\eqref{eq:xz} exhibits state-to-output controllability, a critical property that determines if cognitive perturbations can theoretically steer neural dynamics to any desired cognitive experience. To test this, we utilized an fMRI dataset (detailed in Appendix~\ref{fMRI data}) from five healthy adults (four males, one female, aged 22--34). Each subject viewed 2,185 emotionally evocative short videos, originally curated by Cowen and Keltner~\cite{cowen2017self}, with lengths ranging from 0.15 to 90 seconds (no audio). For each clip \(k\), we obtained: 1) a low-dimensional rating vector, \(\mathbf{u}_k \in \mathbb{R}^{14}\), representing the video's affective attributes, and 2) a corresponding steady-state neural response in voxel space, \(\mathbf{x}_k \in \mathbb{R}^N\).

Using the empirical state-to-output controllability Gramian \(\hat{\mathbf{W}}_z\) (derived in Section~\ref{sec:theory}), we assessed controllability across all five subjects. The Gramians were found to be full rank (i.e., all eigenvalues were nonzero), indicating that, in principle, the system is state-to-output controllable. This means that cognitive perturbations, such as viewing new scenes, could theoretically guide the brain to any desired cognitive state given sufficient time. However, the eigenvalues spanned over two orders of magnitude, with some up to 500 times larger than others. This variability suggests that while full controllability is possible, the effort required differs vastly across dimensions---easy changes align with large eigenvalues, while difficult ones correspond to small eigenvalues. For instance, if one hour of intervention yields a noticeable shift in an ``easy'' direction, the same effort might produce a 500-fold smaller effect in a ``hard'' direction tied to a smaller eigenvalue.

\begin{figure}[t]
    \centering
    \includegraphics[width=\textwidth]{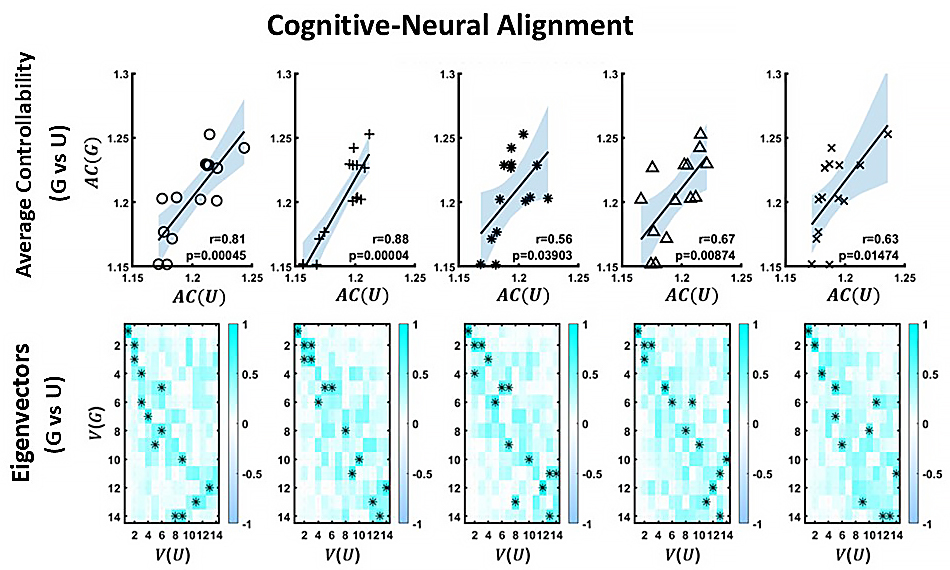}
    \caption{\textbf{Neural--Cognitive Linear Alignment.}
    \textbf{(Top)} Scatter plots of the Average Controllability (AC) at the cognitive level (\(\Pbf_\Ubf\)) vs.\ the neural level (\(\Pbf_{\Gbf^\top}\)) in five subjects. For a given matrix \(\Pbf_{n \times n}\), AC of each dimension \(i\) is a spectral measure quantifying the ease of steering the hypothetical dynamical system \(\xi(t+1) = \Pbf \xi(t) + \mathbf{e}_i \upsilon(t)\) using a hypothetical input \(\upsilon(t)\) which only affects the \(i\)th dimension directly (\(\mathbf{e}_i\) is a binary vector with only one 1 at the \(i\)th location)~\cite{gu2015controllability}. We observe strong positive correlations between ACs of \(\Pbf_\Ubf\) and \(\Pbf_{\Gbf^\top}\) (\(r \approx 0.7\), regression confidence interval shown in blue).
    \textbf{(Bottom)} Heatmaps of cosine similarity between eigenvectors of \(\Pbf_\Ubf\) and \(\Pbf_{\Gbf^\top}\). The diagonal dominance reveals that trajectories defined in the cognitive space are predictably mirrored in the neural state space. Each panel represents data from one subject.}
    \label{fig:Fig3}
\end{figure}

Next, we examined average controllability (AC) to quantify how easily specific dimensions can be influenced. We computed partial correlation matrices for the 14 affective dimensions (the rows of \(\mathbf{U}\)), \(\Pbf_\Ubf\), and for the empirically estimated fMRI gradients (the columns of \(\mathbf{G}\)), \(\Pbf_{\Gbf^\top}\). For a given matrix \(\Pbf_{n \times n}\), the AC of each dimension \(i\) measures the ease of steering a hypothetical system \(\xi(t+1) = \Pbf \xi(t) + \mathbf{e}_i \upsilon(t)\) using an input \(\upsilon(t)\) affecting only the \(i\)th dimension (\(\mathbf{e}_i\) is a binary vector with a single 1 at position \(i\))~\cite{gu2015controllability}. Figure~\ref{fig:Fig3} (top panel) presents scatter plots comparing AC at the cognitive level (\(\Pbf_\Ubf\)) versus the neural level (\(\Pbf_{\Gbf^\top}\)) across the five subjects. The results revealed robust linear correlations (\(r \approx 0.7\)), with regression confidence intervals shown in blue, indicating that dimensions easiest to control cognitively align closely with those easiest to control neurally.

\begin{figure}[t]
    \centering
    \includegraphics[width=1\textwidth, keepaspectratio]{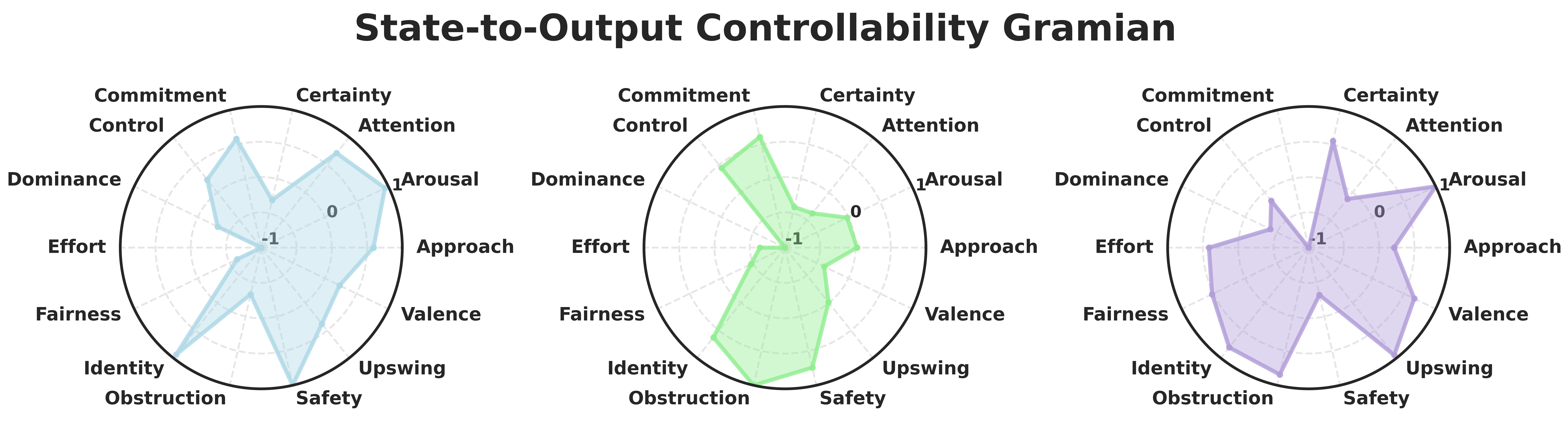}
    \caption{\textbf{State-to-Output Controllability Gramian and Cognitive Trajectories.}
    Radar plots depict the three largest eigenvectors of the controllability Gramian in a 14-dimensional cognitive space. These directions---which combine dimensions such as arousal, valence, control, and approach---are most readily affected by cognitive or psychotherapeutic inputs. Larger magnitudes in an eigenvector indicate greater ease of inducing changes in that dimension. Results are shown across five subjects, whose eigenvectors exhibit a high degree of similarity (mean cosine similarity \(r = 0.6\)).}
    \label{fig:Fig4}
\end{figure}

Finally, we explored the eigenvectors of the controllability Gramian to identify the specific directions of change most amenable to cognitive perturbations. These eigenvectors reveal the principal trajectories along which cognitive experiences evolve, naturally leading to an analysis of cognitive trajectories. Figure~\ref{fig:Fig3} (bottom panel) shows heatmaps of cosine similarity between the eigenvectors of \(\Pbf_\Ubf\) and \(\Pbf_{\Gbf^\top}\), with each row representing an eigenvector of \(\Pbf_{\Gbf^\top}\) and each column an eigenvector of \(\Pbf_\Ubf\). The diagonal dominance (brighter blocks) and statistically significant alignments (asterisks) confirm that the dominant eigenvectors are highly similar across cognitive and neural spaces. Across subjects, the top-ranked eigenvectors exhibited strong pairwise matches (cosine similarity 0.4--0.9, \(p < 10^{-12}\)), suggesting conserved patterns. This led us to Figure~\ref{fig:Fig4}, which illustrates the three largest eigenvectors averaged across subjects (mean \(r = 0.6\)) due to their high similarity. These radar plots highlight combinations of affective dimensions---such as valence, approach, and control---that are most susceptible to alteration via cognitive or psychotherapeutic inputs, with larger magnitudes indicating greater ease of change.

\section{Neural–Cognitive Trajectories and Their Clinical Implications}
\label{sec:validation}

The findings from Section~\ref{sec:fMRIAC} indicate that cognitive and neural dynamics align closely, suggesting a clinical hypothesis: the same low-dimensional “controllability” structure that governs these dynamics may also constrain how psychiatric symptoms evolve in conditions such as Major Depressive Disorder (MDD). In other words, affective state transitions—such as moving from a negative mood to a more positive one through psychotherapy—are likely confined to a limited number of key directions defined by the leading eigenvectors of the controllability Gramian. These directions, in turn, outline the most feasible paths toward clinical improvement.

\begin{figure}[t]
    \centering
    \includegraphics[width=\textwidth, keepaspectratio]{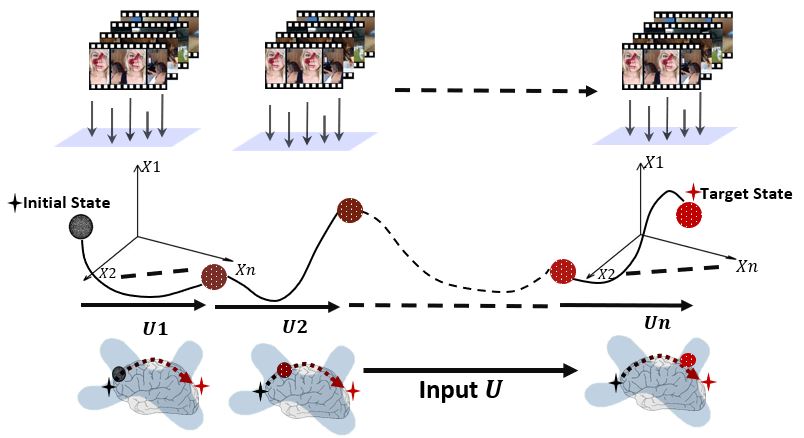}
    \caption{\textbf{Decomposing Extended Cognitive Perturbations into Smaller Inputs.}
    Any extended cognitive intervention (e.g., a long therapy session or repeated stimulus exposures) can be viewed as a sequence of smaller perturbations that each follow the same neural–cognitive dynamics estimated from our original video dataset. In this schematic, the system transitions from an initial state (gray node) toward a target state (red node) via multiple intermediate steps (\(U_1, U_2, \dots, U_n\)). Crucially, the resulting velocity vectors for these steps lie predominantly within the subspace spanned by the eigenvectors of the state-to-output controllability manifold. In Section~\ref{sec:validation}, we demonstrate this principle empirically in a clinical dataset of 255 subjects~\cite{hebbrecht2020understanding}, showing that real-world symptom trajectories in depression likewise adhere to these manifold-based constraints.}
    \label{fig:Fig5}
\end{figure}

To test this hypothesis, we examined symptom data from 255 hospitalized MDD patients, each assessed roughly every two weeks over an 11-week period using the 17-item Hamilton Rating Scale for Depression (HRSD-17; median 5.8 assessments per patient; data originally presented in~\cite{hebbrecht2020understanding}). Our aim was to determine whether the 14-dimensional cognitive–neural subspace derived in Section~\ref{sec:fMRIAC} could explain real-world symptom trajectories. Because the HRSD-17 spans 17 symptom dimensions (e.g., mood, sleep, guilt), whereas our subspace has only 14, we embedded the 14-dimensional directions into the 17-dimensional clinical space by zero-padding and orthonormalizing. Crucially, this embedding did not require matching specific symptom items (e.g., “valence” to “mood”); rather, it tested whether the overall shape of symptom changes aligns with the healthy cognitive–neural patterns.

Next, we quantified how symptoms evolved over time for each patient. Let \(Z \in \mathbb{R}^{T \times 17}\) denote the HRSD-17 scores across \(T\) clinical assessments. For each patient, we computed “velocity vectors,” 
\[
v(t) = Z(t+1, :) - Z(t, :),
\]
capturing symptom change over each two-week interval. We then projected each velocity vector onto the embedded 14-dimensional subspace,
\[
\hat{v}(t) = \bigl(v(t) \, B_{\mathrm{orth}}\bigr) \, B_{\mathrm{orth}}^\top,
\]
and calculated how well these projections matched the observed changes using an \(R^2\) metric:
\[
R^2 = 1 - \frac{\sum_t \|v(t) - \hat{v}(t)\|^2}{\sum_t \|v(t)\|^2}.
\]
This \(R^2\) value represents the proportion of variance in the actual symptom-velocity vectors explained by the subspace.

To ensure the alignment was not due to chance, we performed a permutation test: we randomly shuffled the order of the 17 HRSD-17 items 100 times for each patient, recomputed \(R^2\), and compared the resulting null distribution to the true \(R^2\). The fraction of shuffled \(R^2\) values exceeding the real \(R^2\) served as the \(p\)-value. 

Among the 143 patients with \(T \geq 5\) clinical assessments, 93\% showed an \(R^2\) that exceeded the mean of the shuffled distribution, with over one-third achieving statistical significance (\(p < 0.05\)). These results imply that the geometry of depressive symptom trajectories adheres to the same low-dimensional constraints identified in healthy individuals. In most cases, real symptom changes lay closer to the cognitive–neural manifold than expected by chance, indicating a robust directional preference in how symptoms evolve.

\section{Discussion}
\label{discussion}

The key idea behind our work is that while the cognitive and neural state space is high-dimensional, not all possible patterns are achievable; instead, these trajectories are constrained to a much lower-dimensional manifold that can be quantified and its properties can inform and explain what we observe in human behavior and mental health trajectories. The concept of neural manifolds in high-dimensional brain activity has been studied extensively in computational neuroscience~\cite{mcintosh2019hidden,busch2023multi}, yet their specific implications for cognition and therapy remain underexplored. Our findings suggest that these manifolds serve as geometric scaffolds for cognitive experiences, determining which transitions—such as shifting from a negative bias to a more positive affect—are more feasible under purely cognitive interventions. While we did not impose prior assumptions, our results indicate a nearly one-to-one correspondence between cognitive and neural manifolds, suggesting that the number of dominant controllable dimensions (eigenvectors) within the manifold is relatively limited. This aligns with network control theories that emphasize low-dimensional constraints on brain dynamics~\cite{nozari2024macroscopic}.

Further, our results align with clinical observations and neurostimulation findings, confirming that not all regions of the neural manifold are equally accessible, with some cognitive and emotional trajectories being more feasible than others~\cite{hahn2023towards,muldoon2016stimulation,guidotti2024neuromodulation}. Our analysis of the state-to-output controllability Gramian reveals eigenvalues spanning over two orders of magnitude, indicating that transitions associated with large eigenvalues require minimal effort, while those tied to smaller eigenvalues demand significantly greater energy, time, or therapeutic support. Clinically, this suggests that certain cognitive and affective dimensions are more amenable to intervention, whereas others pose greater challenges. For instance, psychotherapies like cognitive-behavioral therapy (CBT) often effectively target symptoms such as depressed mood and negative thought patterns, facilitating shifts in emotional valence within 10–15 sessions for many patients, yet they frequently struggle with entrenched sleep disturbances and cognitive rigidity~\cite{hofmann2014toward,hayes2021third}. These harder-to-control dimensions often correspond to residual symptoms in Major Depressive Disorder (MDD), such as anhedonia, sleep disturbances, fatigue, guilt, and cognitive rigidity, which persist despite treatment and increase relapse risk~\cite{romera2013residual}. Our findings provide a mathematical framework to elucidate these residual symptoms by mapping them to low-controllability eigenvectors, predicting their persistence using eigenvalue magnitudes, and potentially controlling them by modifying the brain’s dynamic properties. Interestingly, recent evidence from our group and others demonstrates that symptom network dynamics enable accurate prediction of these trajectories, strongly corroborating our controllability-based findings~\cite{stocker2023formalizing,jamalabadi2022complex,hahn2021network,borsboom2017network,scheffer2024dynamical}.

From a therapeutic perspective, identifying dominant controllability directions offers actionable insights: interventions are likely to be most effective when they leverage eigenvectors associated with high controllability, thereby facilitating meaningful cognitive shifts with reduced effort and time. Conversely, more challenging trajectories may require additional interventions, such as neurostimulation or pharmacotherapy, to alter the brain’s underlying controllability properties~\cite{ross2018cost}. This opens a promising avenue for neurostimulation, which could re-engineer neural manifolds beyond traditional state-space control by directly targeting and reshaping the controllability landscape. Such an approach may unlock previously inaccessible regions of the manifold, addressing residual symptoms more effectively, a potential that aligns closely with recent clinical observations~\cite{akhtar2016therapeutic,andrade2010neurostimulatory,morriss2024connectivity}. Incorporating data from neurostimulation studies could further refine this approach, as evidenced by targeted brain state-dependent interventions.

\subsection{Limitations and Future Directions}

\textbf{Temporal and Multiscale Dynamics}. A key limitation of our study is its simplified representation of cognitive and neural dynamics, which does not account for the multiscale interactions observed in real-world brain processes. These interactions are driven by distinct time constants: neural dynamics operate on a fast time scale (\(\tau_x\)), typically on the order of milliseconds to seconds, while cognitive experiences evolve on slower time scales (\(\tau_u\)), often spanning seconds to minutes~\cite{sikka2020investigating, bernacchia2011reservoir, huk2018beyond, berman2016predictability}. To address this, our framework could incorporate a model where cognitive updates occur in discrete blocks of neural steps, potentially altering the controllability landscape. Such a model suggests that larger mismatches between \(\tau_x\) and \(\tau_u\) might complicate the control of less aligned cognitive trajectories, though the precise effects on control dynamics (e.g., time or effort required) remain uncertain and necessitate further theoretical investigation. This omission may bias our controllability Gramian estimates, as multiscale effects could influence the accessibility of certain manifold directions. Multiscale dynamics are a critical factor in mental health, as demonstrated by our recent study on Fourier transformation of Ecological Momentary Assessment (EMA) data~\cite{jamalabadioptimizing}, which reveals their importance in decoding temporal patterns of psychiatric symptoms. Addressing these complexities could enhance the framework’s relevance to real-world mental health applications and may inspire new treatment strategies, such as navigating the controllability manifold to bypass energetically costly directions and target hard-to-control symptom trajectories more effectively.

\textbf{Model Validation and Data Availability}. While our study involving 255 MDD patients underscores the clinical relevance of low-dimensional manifolds, further validation on larger and more diverse datasets is essential. Additionally, many clinical endpoints remain categorical (e.g., “remission” vs. “non-remission”), contrasting with the continuous-state assumptions in our linear model. Future research may benefit from hybrid models that integrate discrete and continuous representations to better capture clinical realities. The inpatient focus of our MDD dataset may introduce bias, limiting generalizability to outpatient populations.

Moreover, our study focused primarily on affective ratings such as arousal and valence, which are central to emotional processing. However, a broader spectrum of cognitive processes—including attentional biases, motivational states, and metacognition—may also be integral to psychiatric disorders \cite{hofmann2014toward,lemoult2019depression}. Expanding the scope of cognitive measures could reveal additional eigenvectors or controllability dimensions relevant to other psychiatric conditions. Lastly, mental health outcomes are shaped by socioeconomic factors, genetic predispositions, environmental stressors, and interpersonal dynamics—variables not explicitly incorporated in our control-theoretic approach. Developing a more holistic framework that integrates these additional influences could enhance predictive accuracy and inform next-generation precision psychiatry.

\section{Conclusion}

Our findings integrate dynamical systems theory, neural-cognitive alignment, and psychiatric intervention research, offering a novel framework that conceptualizes cognitive experience as a state-to-output controllability problem. We provide theoretical and empirical evidence that low-dimensional neural manifolds constrain the cognitive states most readily attainable, with trajectories governed by the controllability Gramian’s eigenvalues. This approach provides a mathematical framework to explain why certain therapeutic interventions succeed while others falter. Our theory and results suggest the potential to quantify and understand human behavior, enabling precise control of cognitive states through targeted interventions. Specifically, combining Ecological Momentary Assessment (EMA) data, neuroimaging, cognitive-behavioral therapy (CBT), and neurostimulation can provide the necessary data and strategies, with further refinement needed to address multiscale dynamics, while dynamical systems and control theory illuminate the path forward.

\backmatter
\section{Declarations}

\subsection{Ethical Approval}

All participants provided written informed consent prior to their involvement in the experiments. The study protocols were reviewed and approved by the respective ethics committees overseeing the datasets utilized in this research, as detailed in Horikawa et al. (2020) and Hebbrecht et al. (2020)~\cite{horikawa2020neural,hebbrecht2020understanding}.

\subsection{Funding Acknowledgment} 
This work was funded in part by the consortia grants from the German Research Foundation (DFG) SFB/TRR 393 (project grant no 521379614),  by a research grant from von Behring Röntgen (No.\ 70\_00038) Stiftung to HJ, by a research grant from the University Hospital of Gießen and Marburg (UKGM, No. 1/2024 MR) to HJ and FB, and by US National Science Foundation Award No. 2239654 to EN.  SGH receives financial support by the Alexander von Humboldt Foundation (as part of the Alexander von Humboldt Professur), the Hessische Ministerium für Wissenschaft und Kunst (as part of the LOEWE Spitzenprofessur), the DYNAMIC center, funded by the LOEWE program of the Hessian Ministry of Science and Arts (Grant Number: LOEWE1/16/519/03/09.001(0009)/98), and the Deutsche Forschungsgemeinschaft (DFG, German Research Foundation) – Project-ID 521379614 – TRR 393SFB/Transregio 393. He also receives compensation for his work as editor from SpringerNature and royalties and payments for his work from various publishers. The funders had no involvement in the study design, data collection, management, analysis, interpretation, report writing, or the decision to submit the report for publication. The authors retained full responsibility for the decision to submit for publication.

\subsection{Data and Code Availability}

All fMRI data used in this study are publicly accessible at \url{https://github.com/KamitaniLab/EmotionVideoNeuralRepresentation}. The clinical data referenced in Section~\ref{sec:validation} were obtained from the authors of Hebbrecht et al. (2020)~\cite{hebbrecht2020understanding} upon reasonable request. All code used to generate the simulations and results will be made publicly available upon publication of this paper at \url{https://www.psycontrol-lab.de/}.  

\subsection{Conflict of Interest}
The Authors declare no competing interests.
\section{Author contribution statement} 
Conceptualization, Methodology, and Validation: HJ, BS, EN. Statistical Analysis: BS, EN, HJ. Data Acquisition: EG, KH, Writing-Original Draft: EN, BS, HJ, Writing-Review: all.



\clearpage
\begin{appendices}

\section{}
\label{Proof_theory1}

\textbf{Detailed derivations on empirical state-to-output controllability Gramian.} In this Appendix We provide simple derivations that better illustrate the theoretical relationship between exact and empirical state-to-output controllability Gramian.

Consider the same state-space model as in~\eqref{eq:xz}, repeated here for convenience:
\begin{subequations}\label{eq:app:xz}
\begin{align}
    \label{eq:app:xz1} \mathbf{x}(t+1) &= \mathbf{A}\,\mathbf{x}(t) + \mathbf{B}\,\mathbf{u}(t), \\
    \mathbf{z}(t)   &= \mathbf{G}^\top\,\mathbf{x}(t).
\end{align}
\end{subequations}
Assume that \( \mathbf{A} \) is stable (i.e., all eigenvalues lie within the unit circle), so that the steady-state response to a constant input $\mathbf{u}_k(t) \equiv \mathbf{u}_k$ exists. In particular, after a sufficiently long time the system reaches
\begin{equation}\label{eq:app:ss_response}
    \mathbf{x}_k = \lim_{t\to\infty} \sum_{\ell=0}^{t-1} \mathbf{A}^\ell \mathbf{B}\,\mathbf{u}_k = \Big[\sum_{\ell=0}^\infty \mathbf{A}^\ell \mathbf{B}\Big] \mathbf{u}_k
    = (\mathbf{I}-\mathbf{A})^{-1}\mathbf{B}\,\mathbf{u}_k.
\end{equation}
The last expression also follows directly from~\eqref{eq:app:xz1} as its fixed point solution--setting $\xbf(t+1) = \Xbf(t)$.
Now suppose we collect steady-state responses $\xbf_k, k = 1, \dots, K$ for \( K \) different cognitive stimuli $\ubf_k, k = 1, \dots, K$ and define the data matrices
\[
\mathbf{X} = \begin{bmatrix} \mathbf{x}_1 & \mathbf{x}_2 & \cdots & \mathbf{x}_K \end{bmatrix} \quad\text{and}\quad
\mathbf{U} = \begin{bmatrix} \mathbf{u}_1 & \mathbf{u}_2 & \cdots & \mathbf{u}_K \end{bmatrix}.
\]
Then, from~\eqref{eq:app:ss_response}, we have
\begin{align}\label{eq:XIABU}
\mathbf{X} = (\mathbf{I}-\mathbf{A})^{-1}\mathbf{B}\,\mathbf{U}.
\end{align}
If \( \mathbf{U} \in \mathbb{R}^{n \times K} \) is full row rank (a reasonable assumption when \( K \gg n \)), the unique solution to the system of linear equation~\eqref{eq:XIABU} is
\begin{equation}\label{eq:app:lifted}
    (\mathbf{I}-\mathbf{A})^{-1}\mathbf{B} = \mathbf{X}\,\mathbf{U}^\dagger.
\end{equation}
This can then be used to approximate the state-to-output controllability Gramian, as follows.
The state-to-output controllability Gramian is defined as
\begin{align}\label{eq:app:Wz_true}
    \notag \mathbf{W}_z = \sum_{\ell=0}^{\infty} \mathbf{G}^\top \mathbf{A}^\ell \mathbf{B}\,\mathbf{B}^\top (\mathbf{A}^\top)^\ell \mathbf{G}
    &= \Gbf^\top \Big[\sum_{\ell=0}^{\infty} \mathbf{A}^\ell \mathbf{B}\,\mathbf{B}^\top (\mathbf{A}^\top)^\ell \Big] \Gbf
    \\
    \notag &\simeq \Gbf^\top \Big[\sum_{\ell=0}^{\infty} \mathbf{A}^\ell \mathbf{B} \Big] \Big[\sum_{\ell=0}^{\infty} \mathbf{B}^\top (\mathbf{A}^\top)^\ell \Big] \Gbf
    \\
    &= \mathbf{G}^\top (\mathbf{I}-\mathbf{A})^{-1}\mathbf{B}\,\mathbf{B}^\top (\mathbf{I}-\mathbf{A})^{-T} \mathbf{G} = \hat \Wbf_z.
\end{align}
The last equality serves as the definition of the empirical Gramian $\hat \Wbf_z$. Note that $\hat \Wbf_z$ is only an approximation of $\Wbf_z$ because the product of the two infinite series on the second line of~\eqref{eq:app:Wz_true} has extra terms in addition to the terms in the preceding infinite series. Using the relation in~\eqref{eq:app:lifted}, we then have
\begin{equation}\label{eq:app:Wz_empirical}
    \hat{\mathbf{W}}_z = \mathbf{G}^\top \left(\mathbf{X}\,\mathbf{U}^\dagger\right)
    \left(\mathbf{X}\,\mathbf{U}^\dagger\right)^\top \mathbf{G}.
\end{equation}
Since the matrices \( \mathbf{X} \) and \( \mathbf{U} \) are computed from the steady-state responses, 
one can compute $\hat \Wbf_z$ and its spectrum directly from data, thereby assessing the energetic cost of steering the output \( \mathbf{z}(t) \).

\section{}\label{Proof_theory2}

\textbf{Detailed derivations on the linearity of fMRI gradients' relationship to cognitive perturbations.} Here we demonstrate that, under a linear dynamical model, the structure of the fMRI gradients is linearly related to the correlations among cognitive inputs.
Assuming that the range of each component of the cognitive input \( \mathbf{u} \) is normalized to lie in \([0,1]\), the fMRI gradient associated with the \( i \)-th cognitive experience is
\begin{equation}\label{eq:app:gi_def}
    \gbf_i = \mathbb{E}[\mathbf{x} \mid u_i = 1] - \mathbb{E}[\mathbf{x} \mid u_i = 0].
\end{equation}
Using the steady-state response from~\eqref{eq:app:ss_response}, we have
\[
\mathbb{E}[\mathbf{x} \mid u_i = c] = (\mathbf{I}-\mathbf{A})^{-1}\mathbf{B}\,\mathbb{E}[\mathbf{u} \mid u_i = c], \quad c \in \{0,1\}.
\]
Hence,
\begin{equation}\label{eq:app:gi_2}
    \gbf_i = (\mathbf{I}-\mathbf{A})^{-1}\mathbf{B} \left( \mathbb{E}[\mathbf{u} \mid u_i = 1] - \mathbb{E}[\mathbf{u} \mid u_i = 0] \right).
\end{equation}
Assuming that components of $\ubf$ are approximately related to each other via a linear stochastic relationship, we have
\[
\mathbb{E}[u_j \mid u_i = 1] = \rho_{ji} \quad\text{and} \quad \mathbb{E}[u_j \mid u_i = 0] = 0,
\]
where \( \rho_{ji} \) is the correlation coefficient between \( u_j \) and \( u_i \) (with \( \rho_{ii}=1 \)). Denote by \( \rhob_i \) the \( i \)-th column of the cognitive perturbation correlation matrix \( \Pbf \); that is, \( \rhob_i = [\rho_{1i},\,\rho_{2i},\,\ldots,\,\rho_{ni}]^\top \). Then,~\eqref{eq:app:gi_2} becomes
\begin{equation}\label{eq:app:gi_final}
    \gbf_i = (\mathbf{I}-\mathbf{A})^{-1}\mathbf{B}\,\rhob_i.
\end{equation}
This shows that under the linear model in~\eqref{eq:app:xz} the fMRI gradients are linearly related to correlation coefficients between the components of cognitive perturbations.

Next, consider the inner product between two fMRI gradients \( \gbf_i \) and \( \gbf_j \):
\[
\gbf_i^\top \gbf_j = \rhob_i^\top \left[(\mathbf{I}-\mathbf{A})^{-T}\mathbf{B}^\top \mathbf{B} (\mathbf{I}-\mathbf{A})^{-1}\right] \rhob_j.
\]
Therefore, defining the symmetric matrix
\[
\Qbf = \mathbf{B}^\top (\mathbf{I}-\mathbf{A})^{-T} (\mathbf{I}-\mathbf{A})^{-1} \mathbf{B},
\]
we have
\begin{equation}\label{eq:app:inner_prod}
    \gbf_i^\top \gbf_j = \rhob_i^\top \,\Qbf\, \rhob_j.
\end{equation}
This expression shows that the inner product (and hence the correlation) between \( \gbf_i \) and \( \gbf_j \) is determined by a quadratic form in the corresponding vectors \( \rhob_i \) and \( \rhob_j \) of the cognitive correlation matrix \( \Pbf \), with weighting given by \( \Qbf \). Therefore, the matrix of all pairwise correlations among the fMRI gradients is, approximately up to a scaling factor, given by
\[
\Pbf \Qbf \Pbf.
\]
This clearly demonstrates that if the brain dynamics are well approximated by a linear model, then the (partial) correlations among the fMRI gradients mirror the linear structure of the cognitive inputs.

\section{}
\label{fMRI data}

\textbf{fMRI data and processing.} We used an fMRI dataset originally collected and made available by Cowen and Keltner~\cite{cowen2017self}. Five healthy adult participants (four males, one female, aged 22--34) viewed a total of 2{,}185 unique emotionally evocative short videos. These videos were extracted from an original set of 2{,}196 clips whose durations ranged from approximately 0.15\,s to 90\,s. Each video was shown without sound, resized to fit within a 12\textdegree{} visual angle, and presented against a gray background. Participants underwent multiple fMRI scanning sessions over roughly two months, yielding approximately eight hours of data. All procedures for video presentation, participant head stabilization, and run structuring followed the protocols described in the original study. Rest periods were interspersed between video blocks (each 7--10\,min run contained 36~video blocks, plus a 2\,s rest following each block).

Functional images were acquired using a 3.0\,T Siemens MAGNETOM Verio scanner (TR\,=\,2000\,ms, TE\,=\,43\,ms, flip angle\,=\,80\textdegree, 2\,$\times$\,2\,$\times$\,2\,mm voxels, 76 slices, multiband factor\,=\,4). T1-weighted anatomical scans were also collected. Preprocessing was conducted with \texttt{fMRIPrep}~\cite{esteban2019fmriprep}.
After preprocessing, nuisance regression was performed (polynomial trends and motion parameters). The time series were shifted by 4\,s to account for hemodynamic delay and despiked to remove extreme values. We then averaged voxel signals within each stimulus block (i.e., a video's on-period plus a 2\,s rest), and z-scored across all stimuli per voxel. This yielded one fMRI response vector per video, per subject. We used whole brain data these data for subsequent neural--cognitive alignment and controllability analyses.

Each of the 2{,}185 video stimuli was annotated with a 14-dimensional affective rating, provided by an online crowd-sourced protocol (see Cowen and Keltner~\cite{cowen2017self} for details). Participants in that separate study rated the videos along 14 affective dimensions (e.g., arousal, valence) on a 9-point scale, and ratings were averaged across raters. These 14 scores are treated as low-dimensional descriptors of each video's emotional content.

\end{appendices}

\end{document}